\documentclass[twocolumn,epjc3]{svjour3}

\usepackage{graphicx,amssymb}
\usepackage{hyperref}

\def \d {{\rm d}}
\def \e {e}

\begin{document}

\title{Kundt spacetimes minimally coupled to scalar field}

\author{T. Tahamtan\thanksref{e1,addr1,addr2}
        \and
        O. Sv\'{\i}tek\thanksref{e2,addr1}}
\thankstext{e1}{e-mail: tahamtan@utf.mff.cuni.cz}
\thankstext{e2}{e-mail: ota@matfyz.cz}
\institute{Institute of Theoretical Physics, Faculty of Mathematics and Physics, Charles University, V~Hole\v{s}ovi\v{c}k\'ach 2, 180~00 Prague~8, Czech Republic\label{addr1}
\and
 Astronomical Institute, Czech Academy of Sciences, Bo\v{c}n\'{\i} II 1401, Prague, CZ-141 31, Czech Republic\label{addr2}}

\date{\today}

\maketitle

\begin{abstract}
	We derive an exact solution belonging to Kundt class of spacetimes both with and without a cosmological constant that are minimally coupled to a free massless scalar field. We show the algebraic type of these solutions and give interpretation of the results. Subsequently, we look for solutions additionally containing an electromagnetic field satisfying nonlinear field equations.

\end{abstract}

\section{Introduction}

\sloppy

The Kundt class of spacetimes is among the most important families of exact solutions of Einstein equations. It was derived more that half a century ago \cite{Kundt1,Kundt2} and it is defined by the presence of a nonexpanding, nonshearing and nontwisting null geodesic congruence \cite{Stephanietal:book,GriffithsPodolsky:book}. This class contains solutions of various algebraic types and admits, apart from vacuum solutions, a cosmological constant, an electromagnetic field, a gyraton source or solutions with supersymmetry. All type D vacuum solutions were classified in the classic paper \cite{Kinnersley}. This family also contains the famous pp-wave solution \cite{Stephanietal:book,GriffithsPodolsky:book} and so called VSI (vanishing scalar invariants) \cite{VSI} or CSI (constant scalar invariants) \cite{CSI} spacetimes. The Kundt class of spacetimes was also recently generalized to arbitrary dimension \cite{PodolskyZofka:2009}. Subclasses of the Kundt family of spacetimes provide the most important examples of so--called universal spacetimes---solutions of vacuum field equations of arbitrary gravitational theory whose Lagrangian is a polynomial invariant constructed from the metric, the Riemann tensor and its derivatives of arbitrary order \cite{Hervik}. Universal metrics also represent classical solutions to string theory because associated quantum corrections vanish in this case \cite{Coley}.

Solutions to Einstein equations containing scalar field serve as a useful tool for understanding General Relativity due to the simplicity of the source. Recently, it has become evident that fields of this type really do exist (LHC) and play a fundamental role in the Standard model of particle physics. In classical General Relativity they were used to study counterexamples to black hole no-hair theorems or cosmic censorship hypothesis and in many other areas. The study of scalar fields in Kundt spacetimes complements the one performed in the closely related Robinson--Trautman family \cite{Tahamtan,Tahamtan2}.

Kundt spacetimes with Maxwell electrodynamics were extensively analyzed soon after the appearance of vacuum solutions. One of the most important examples is the conformally flat Bertotti--Robinson solution \cite{Bertotti-Robinson} which contains a uniform non-null electromagnetic field. More general solutions possibly containing additional pure radiation and (exact) gravitational waves were found as well (for a review see, e.g., \cite{Ortaggio-Podolsky}). The pure radiation solutions can be used to support perturbative gravitational waves as well \cite{Podolsky-Svitek-GRG}.

Nonlinear Electrodynamics (NE) was originally used mainly as a solution to the problem of divergent field of a point charge in the vicinity of its position (see e.g. \cite{Dirac}) also giving satisfactory self-energy of charged particle. The best-known and frequently used form of the theory was introduced already in 1934 by Born and Infeld \cite{BornInfeld}. A nice overview with a lot of useful information was given in a book by Pleba\'{n}ski \cite{Plebanski}. Apart from solving the point charge singularity the NE was later used to resolve the spacetime singularity as well.

In this work, we first give explicit solutions of Kundt type with minimally coupled free scalar field. So far the interest was mainly directed towards analyzing pp-waves coupled with scalar field and possibly additional sources (e.g. Yang-Mills fields \cite{Balakin}) which stemmed from the extensive use of pp-waves in string theory. In the next part, we concentrate on solutions containing nonlinear electromagnetic field as an additional source, complementing a similar study done in the Robinson--Trautman family \cite{Tahamtan3}.

\fussy

\section{Field equations with scalar field}\label{first section}
We consider the following action, describing a free massless scalar field minimally coupled to a gravity described by the Einstein-Hilbert action 
\begin{equation}\label{action}
S=\frac{1}{2}\int d^{4}x \sqrt{-g}[R+\nabla_{\mu}\varphi \nabla^{\mu} \varphi-2\Lambda],
\end{equation}
where R is the Ricci scalar for the metric $g_{\mu \nu}$, and we have included cosmological constant $\Lambda$. The massless scalar field (SF) $ \varphi $ is considered to be real and we use units in which $c=\hbar=8 \pi G=1$.

By applying variation with respect to the metric we get the following field equations for the action (\ref{action}) 
\begin{equation}\label{field equations}
G^{\mu}{}_{\nu}={}^{\rm SF}T^{\mu}{}_{\nu}-\Lambda \delta ^{\mu}{}_{\nu}\,,
\end{equation}
where the energy momentum tensor of the scalar field is given by
\begin{equation} \label{energy-momentum scalar-field}
{}^{\rm SF}T_{\mu \nu}=\nabla_{\mu}\varphi\, \nabla_{\nu}\varphi-\frac{1}{2}g_{\mu \nu}\,g^{\alpha\beta}\nabla_{\alpha}\varphi \nabla_{\beta}\varphi\,.
\end{equation}

One can easily check that the full energy momentum tensor on the r.h.s. of equation (\ref{field equations}) (as well as its scalar field and cosmological constant parts individually) satisfies the null energy condition ($T_{\mu\nu}l^{\mu}l^{\nu}\geq 0$ for any $\l^{\mu}l_{\mu}=0$). Then the Kundt class of spacetimes defined in \cite{Stephanietal:book} as nonexpanding and nontwisting is necessarily also nonshearing for such a matter content and we can assume the following form of the general line element for the Kundt spacetime (derived from properties of the geometrically preferred null congruence, see \cite{Stephanietal:book,GriffithsPodolsky:book}) which is suitable for analyzing all the explicit solutions presented below 
\begin{eqnarray}\label{ourmetric}
 \d s^2&=&-H\,\d u^2-2\,\d u\d v -2\,W_{1}\,\d u\d x - 2\,W_{2}\,\d u\d y+\nonumber\\
 &&\qquad +\frac{\d x^2+\d y^2}{P(u,x,y)^2}\,,
\end{eqnarray}
with $H, W_{1}, W_{2}$ being functions of all coordinates. The coordinate $v$ is an affine parameter along the principal null congruence $\partial_{v}$, hypersurfaces $u=const$ are null and coordinates $x,y$ span the transversal spatial 2-space.

The following coordinate transformations together with redefinitions of the metric functions $H$, $W_{1}$, $W_{2}$ and $P$ are preserving the form of the metric (\ref{ourmetric}) \cite{Stephanietal:book},
\begin{itemize}
\item[i)]
\begin{eqnarray}
&&\,v'=v+g(u,x,y),\,\,\,P'=P,\,\,\,\,\,\,\,\,H'=H-g_{,u} \nonumber\\
&&W'_{1}=W_{1}-g_{,x}\,,\,\,\,\,\,\,\,\,\,W'_{2}=W_{2}-g_{,y}
\end{eqnarray}
\item[ii)]
\begin{eqnarray}
&&\,u'=h(u),\,\,\,\,\,\,\,v'=\frac{v}{h_{,u}}\,,\,\,\,H'={(H+v\frac{h_{,uu}}{h_{,u}})}/{(h_{,u})^2} \nonumber\\
&&P'=P,\,\,\,\,\,\,\,\,\,\,\,\,\,\,W'_{1}=\frac{W_{1}}{h_{,u}}\,,\,\,\,\,\,\,\,\,\,\,\,W'_{2}=\frac{W_{2}}{h_{,u}}
\end{eqnarray}
\end{itemize}
additionally it is possible to perform a $u$-dependent transformation in the transversal space spanned by coordinates $x,y$ followed by appropriate metric function changes (check \cite{Stephanietal:book} for details).

The Ricci scalar for the metric (\ref{ourmetric}) is 
\begin{eqnarray}
	\mathcal{R}&=&2\Delta(\,\ln P)+2P^2\,(W_{1,xv}+W_{2,yv})\\
	&&-\frac{3}{2}P^2\,({W_{1,v}}^2+{W_{2,v}}^2)\nonumber\\
	&&-2P^2\,(W_{1}W_{1,vv}+W_{2}W_{2,vv})-H_{,vv}\nonumber\,,
\end{eqnarray}
where 
\[\Delta\equiv P(u,x,y)^2(\partial_{xx}+\partial_{yy})\ .\]
 By computing optical scalars of the congruence generated by null geodesic vector $\partial_{v}$ one indeed obtains vanishing expansion, shear and twist.

Now we proceed to the simplification of metric functions and scalar field arising from the Einstein equations. Initially, we consider the scalar field being a function of all the coordinates $\varphi(u,v,x,y)$. The following equation provides straightforward restrictions on the dependence of the scalar field in the $v$ direction
\begin{equation}\label{v-constraint}
	G^u{}_v=0=(\varphi_{,v})^{2}={}^{\rm SF}T^{u}{}_{v}\,,
\end{equation}
so that from now on we can consider $\varphi(u,x,y)$ as the most general admissible form of the scalar field (see Appendix for further discussion of the complex scalar field case). This means that certain off-diagonal energy momentum tensor components are vanishing now which provides restrictions on metric functions through the Einstein equations
\begin{eqnarray}
	G^x{}_v=\frac{1}{2}P^{2}\,W_{1,vv}&=&0={}^{\rm SF}T^{x}{}_{v}\,,\\
	G^y{}_v=\frac{1}{2}P^{2}\,W_{2,vv}&=&0={}^{\rm SF}T^{y}{}_{v}\,,\\
	G^u{}_x=-\frac{1}{2}\,W_{1,vv}&=&0={}^{\rm SF}T^{u}{}_{x}\,,\\
	G^u{}_y=-\frac{1}{2}\,W_{2,vv}&=&0={}^{\rm SF}T^{u}{}_{y}\,.
\end{eqnarray}
Evidently, the functions $W_{1}, W_{2}$ are only linear in $v$.

The scalar field must satisfy the corresponding field equation
\begin{equation}\label{box}
\Box \varphi=P^{2}({\varphi_{,xx}}+{\varphi_{,yy}}-W_{1,v}\, \varphi_{,x}-W_{2,v}\, \varphi_{,y})=0\,,
\end{equation}
where $\Box$ is a standard d'Alembert operator for our metric (\ref{ourmetric}). Because of the linearity of $W_{1}, W_{2}$ the scalar field wave equation contains only terms independent of $v$. We will specifically consider
\[W_{1}(u,v,x,y)=v\, V_{1}(u,x,y),\] 
\[W_{2}(u,v,x,y)=v\, V_{2}(u,x,y),\] 
since the potential addition of a term independent of $v$ does not bring any substantial change and follows the logic of Theorem 31.1 from \cite{Stephanietal:book}. If this additional term creates a new contribution to the energy momentum tensor components $T_{ux}, T_{uv}$ and generates certain nonvanishing contour integral in the transversal 2-space it is possible to interpret it as a so-called gyraton \cite{Krtous} (gravitational field created by a source with internal rotation).

Since the metric (\ref{ourmetric}) is still quite general in the following we will simplify its structure further using the remaining Einstein equations. The form of $uu$, $xx$ and $yy$ components of the Einstein tensor for the metric (\ref{ourmetric}) is (with $G^u{}_u=G^v{}_v$)
\begin{eqnarray}
	G^u{}_u&=&-\Delta(\,\ln P)+\frac{P^2}{4}\{{V_{1}}^2+{V_{2}}^2-2(V_{1,x}+V_{2,y})\}\,,\nonumber\\
	G^x{}_x&=&\frac{P^2}{4}\{{V_{1}}^2+3{V_{2}}^2\}-(PV_{2})_{,y}P+PV_{1}P_{,x}+\frac{H_{,vv}}{2}\,,\nonumber\\
	G^y{}_y&=&\frac{P^2}{4}\{3{V_{1}}^2+{V_{2}}^2\}-(PV_{1})_{,x}P+PV_{2}P_{,y}+\frac{H_{,vv}}{2}\,.\nonumber
\end{eqnarray}
From field equations (\ref{field equations}) for $xx$ and $yy$ components we obtain
\begin{eqnarray}
xx:&&\frac{P^2}{4}\{{V_{1}}^2+3{V_{2}}^2\}-\frac{P^2}{2}\{{\varphi_{,x}}^2-{\varphi_{,y}}^2\}- (PV_{2})_{,y}P \nonumber \\
&&+PV_{1}P_{,x}+\frac{H_{,vv}}{2}+\Lambda=0\,,\label{xx-comp}\\ 
yy:&&\frac{P^2}{4}\{3{V_{1}}^2+{V_{2}}^2\}+\frac{P^2}{2}\{{\varphi_{,x}}^2-{\varphi_{,y}}^2\}-(PV_{1})_{,x}P \nonumber \\
&&+PV_{2}P_{,y}+\frac{H_{,vv}}{2}+\Lambda=0 \,,
\end{eqnarray}
and by adding these two equations together we have
\begin{equation}\label{xx and yy}
P^2\,({V_{1}}^2+{V_{2}}^2-V_{1,x}-V_{2,y})+H_{,vv}+2\, \Lambda=0\ .
\end{equation}

\sloppy

Since all the terms in the equation (\ref{xx and yy}) except $H_{,vv}$ are independent of $v$ the function $H$ is necessarily a quadratic function in $v$. The standard form of the Kundt metric with vanishing spin coefficient $\tau$ (which means that the privileged null congruence is recurrent \cite{Pravda}) possesses quadratic term of $H$ which is completely determined by the cosmological constant and we will keep this for our $H$ as well

\fussy
\begin{equation} \label{definition of H}
H(u,v,x,y)=-\Lambda\, v^2+M(u,x,y)v+K(u,x,y)\,,
\end{equation}
with $M$ and $K$ arbitrary functions. So now the equation (\ref{xx and yy}) simplifies into 
\begin{equation}\label{v and v_x}
{V_{1}}^2+{V_{2}}^2=V_{1,x}+V_{2,y}\,.
\end{equation}
 Another important equation comes from  $uu$ component of the field equations
\begin{eqnarray}
uu&:&\frac{P^2}{4}\{{V_{1}}^2+{V_{2}}^2-2(V_{1,x}+V_{2,y})+2({\varphi_{,x}}^2+{\varphi_{,y}}^2)\}\nonumber\\
&&-\Delta(\,\ln P)+\Lambda=0\,.\label{uu-comp}
\end{eqnarray}
Using (\ref{v and v_x}) in the above equation we can simplify it into the following form 
\begin{equation}\label{uu equation}
-\Delta(\,\ln P)+\Lambda=\frac{P^2}{4}\{{V_{1}}^2+{V_{2}}^2-2({\varphi_{,x}}^2+{\varphi_{,y}}^2)\}\,.
\end{equation}
The expression $\Delta(\,\ln P)$ is the Gaussian curvature of the two-space of constant $u,v$ equipped with an induced metric coming from (\ref{ourmetric}). In the geometrically simple case of two-surfaces with constant Gaussian curvature which is completely determined by a cosmological constant we have zero on the left-hand side of (\ref{uu equation})). We will study this special case in detail in section \ref{previous}. In the next section we consider the case of two-spaces with non-constant Gaussian curvature determined by (\ref{uu equation})).

Further simplification of our problem arises by considering the following components of Einstein equations
\begin{eqnarray}
xu:&&\left[{P^2}\left(V_{2}\,V_{2,x}-V_{2}\,V_{1,y}+\frac{V_{1,yy}-V_{2,xy}}{2}\right)\right.\nonumber \\
&&+P\,P_{y}\,(V_{1,y}-V_{2,x})+V_{1}\Lambda \bigg]v+\left(\frac{P_{,u}}{P}\right)_{,x}+\frac{M_{,x}}{2} \nonumber\\
&&+V_{1}\, \frac{P_{,u}}{P}-\frac{V_{1,u}}{2}-\varphi_{,u}\, \varphi_{,x}=0\label{xu}\,,\\
yu:&&\left[{P^2}\left(V_{1}\,V_{1,y}-V_{1}\,V_{2,x}-\frac{V_{2,xx}-V_{1,xy}}{2}\right)\right.\nonumber\\
&&+P\,P_{x}\,(V_{2,x}-V_{1,y})+V_{2}\Lambda \bigg]v-\left(\frac{P_{,u}}{P}\right)_{,y}+\frac{M_{,y}}{2}\nonumber\\ 
&&-V_{2}\, \frac{P_{,u}}{P}+\frac{V_{2,u}}{2}+\varphi_{,u}\, \varphi_{,y}=0\label{yu}\,.
\end{eqnarray}

Here again the terms with different powers of coordinate $v$ need to vanish separately. Before proceeding further we impose certain assumptions regarding the form of metric functions and the scalar field. Specifically, we will use the following separation of variables for the scalar field $\varphi$ and functions $M$ and $P$
\begin{eqnarray}
M(u,x,y)&=&h(u)\,{\tilde{M}(x,y)}+\mu(u)\,,\nonumber\\
\varphi(u,x,y)&=&\phi(u)+\psi(x,y)\label{separation} \,,\\
P(u,x,y)&=&\frac{\tilde{P}(x,y)}{U(u)}\,.\nonumber
\end{eqnarray}

\sloppy
By substituting these assumptions into (\ref{box}) we obtain the factorization of the dependence on coordinate $u$ for functions $V_{1}=f(u)\tilde{V}_{1}(x,y)+\bar{V}_{1}(x,y), V_{2}=f(u)\tilde{V}_{2}(x,y)+\bar{V}_{2}(x,y)$ with $\frac{\tilde{V}_{1}}{\tilde{V}_{2}}=-\frac{\psi_{,y}}{\psi_{,x}}$. Further using equation (\ref{v and v_x}) one concludes that necessarily $f= const$. So from now on $V_{1}, V_{2}$ are functions of $x,y$ only. After substituting the above definitions (\ref{separation}) into (\ref{xu}, \ref{yu}) we obtain the following relations when considering the zero order terms in $v$
\begin{eqnarray}
   -\frac{h}{2}\tilde{M}_{,x}+V_{1}\,{(\ln{U})_{,u}}+{\psi_{,x}}{\phi_{,u}}=0\,, \nonumber \\
    -\frac{h}{2}\tilde{M}_{,y}+V_{2}\,{(\ln{U})_{,u}}+{\psi_{,y}}{\phi_{,u}}=0 \,.\label{split-eq}
\end{eqnarray}
\fussy

If we take a derivative of the first equation of (\ref{split-eq}) with respect to $y$ and of the second with respect to $x$, we arrive at the following condition
\begin{equation}\label{V1yV2x}
	V_{1,y}=V_{2,x}\,,
\end{equation}
If we plug this expression back into (\ref{xu}, \ref{yu}) and consider the first order terms in $v$ we immediately conclude that $\Lambda$ has to vanish unless $V_{1}=0=V_{2}$. In the following we split the investigation accordingly.

\section{Singular model}
Here we consider solutions which have $V_{1}, V_{2}$ generally nonvanishing and $\Lambda=0$. Then the Gaussian curvature of the two-surfaces spanned by coordinates $x,y$ is not a constant in general (see (\ref{uu equation})).

We consider the equations (\ref{split-eq}) again. Since all the terms of these equations are of the separated form there are only two possibilities how to satisfy them. Either one factorizes the $x,y$ dependence completely and is left with a simple equation for functions of $u$. Or one proceeds inversely --- factorizing the $u$ dependence out and obtaining an equation for functions of $x,y$. We will follow the less restrictive first option which leads (up to  trivial multiplicative constants) to the following conditions
\begin{eqnarray}
V_{1}(x,y)&=&-\frac{1}{2}\frac{\partial \tilde{M}(x,y)}{\partial x}=\frac{\partial \psi(x,y)}{\partial x} \,,\nonumber \\
V_{2}(x,y)&=&-\frac{1}{2}\frac{\partial \tilde{M}(x,y)}{\partial y}=\frac{\partial \psi(x,y)}{\partial y}\,, \label{V1-condition}
\end{eqnarray} 
and the equation 
\begin{equation}\label{u-dependence}
	h+{(\ln{U})_{,u}}+{\phi_{,u}}=0\,.
\end{equation}

Note that by combining (\ref{V1-condition}) and the previously derived $\Lambda=0$ with (\ref{uu equation}) we conclude that ${ \Delta\,\ln {P} } \geq 0$. This means that the transversal surfaces have necessarily a non-negative Gaussian curvature.

From (\ref{box}), one can already find the scalar field explicitly now
\begin{equation}
\varphi(u,x,y)=\phi(u)-\ln(a+\ln(x^2+y^2))\,,
\end{equation} 
where $ a $ is an arbitrary constant. The equation (\ref{uu equation}) simplifies into
\begin{equation}\label{equationP}
{ \Delta(\,\ln \tilde{P}) }=\frac{1}{4}\Delta\psi\,,
\end{equation}
which can be now solved for $\tilde{P}$ (including homogeneous solution contribution)
\begin{equation}\label{functionP}
\tilde{P}(x,y)=\frac{\sqrt{x^2+y^2}}{\left[a+\ln(x^2+y^2)\right]^{1/4}}\,\,.
\end{equation}

The final nontrivial equation which was not mentioned yet is the $vu$ component of the Einstein equations (\ref{field equations}) which now simplifies into the following two equations representing the zero and  the first order terms in $v$
\begin{eqnarray}
&&\Delta K +K\Delta\psi + {\tilde{P}^2}\left[K_{,x}\,\psi_{,x}+K_{,y}\,\psi_{,y}\right]-2\,\tilde{M}h\,U_{,u}\,U\nonumber\\
&&-2\,\mu\,U_{,u}\,U-4U\,U_{,uu}-2U^2\,{\phi_{,u}}^2=0\,, \label{equation vu}\\
&&h+{(\ln{U})_{,u}}+{\phi_{,u}}=0\,.
\end{eqnarray}
We can see that the second equation is consistent with previously derived condition (\ref{u-dependence}). We will use the separation of variables once more for the function $K$
\[K(u,x,y)=k(u)\,\tilde{K}(x,y)\ .\]
Considering the coordinate dependence for the terms of the equation (\ref{equation vu}) we get a necessary condition for its solvability
\[k=h\,U\,U_{,u}\,\,,\]
and the equation (\ref{equation vu}) splits into a condition for function $k(u)$
\begin{equation}
k(u)=C_{0}(2\,\mu\,U_{,u}\,U+4U\,U_{,uu}+2U^2\,{\phi_{,u}}^2)\,, 
\end{equation}
where $ C_{0} $ is an arbitrary constant and a linear second order elliptic PDE with known coefficients whose solution always exists.

Now, let us determine the algebraic type of important tensors characterizing our solution. We would like our solution to be of a reasonable generality which can be checked based on Petrov or Segre classification. We will use the following tetrad ($i$ is an imaginary unit)
\begin{eqnarray}
  \mathbf{l}&=&\partial_{u}-\left[\frac{H}{2}+\frac{v^{2}\tilde{P}^{2}}{2U^{2}}(V_{1}^{2}+V_{2}^{2})\right]\partial_{v}+\nonumber\\
  &&+\frac{v\tilde{P}^{2}}{U^{2}}(V_{1}\partial_{x}+V_{2}\partial_{y})\,,\nonumber\\
  \mathbf{n}&=&\partial_{v}\,,\\
  \mathbf{m}&=&\frac{\tilde{P}}{\sqrt{2}U}(\partial_{x}-i\partial_{y})\,. \nonumber
\end{eqnarray}
There are three nonvanishing Weyl scalars in this frame
\begin{eqnarray} \label{Weyl-singular}
	\Psi_0&=&\frac{\tilde{P^2}}{4\,U^2} \left\{ K_{,yy}-K_{,xx}+2\,i\,K_{,xy}\right\} \nonumber \\
	&&	+\left(\frac{2\zeta-3}{4\,U^2\,\zeta^{\frac{3}{2}}}\right)\,{(K_{,y}+i\,K_{,x})(y+i\,x)} \\
&&+\left(\frac{3K-10v\,h}{2U^2\,\zeta^{\frac{5}{2}}}+\frac{2v^2}{U^4\,\zeta^{5}}\right)\,\frac{(y+i\,x)^2}{y^2+x^2}\,, \nonumber\\
\Psi_1&=&\frac{\left(v-h\,U^2\,\zeta^{\frac{5}{2}}\right)}{\sqrt{2}\,U^3\zeta^{\frac{15}{4}}}\frac{\left(x-i\,y\right)}{\sqrt{y^2+x^2}}\,,\\
\Psi_2&=&-\frac{1}{6\,{\zeta}^\frac{5}{2}\,U^2}\,,
\end{eqnarray}
where
\begin{equation}\label{singularity}
  \zeta=a+\ln{(x^2+y^2)}\,.
\end{equation}
We use the classification process described in \cite{Zakhary} which is based on \cite{Penrose} and can be used for arbitrary tetrad. Computing the invariants
$$I=\Psi_{0}\Psi_{4}-4\Psi_{1}\Psi_{3}+3\Psi_{2}^{2}\,,\ J={\rm det}\left(\begin{array}{ccc}
\Psi_{4} & \Psi_{3} & \Psi_{2}\\
\Psi_{3} & \Psi_{2} & \Psi_{1}\\
\Psi_{2} & \Psi_{1} & \Psi_{0}
\end{array}\right)$$
one can immediately confirm that $I^{3}=27J^{2}$ is satisfied so that we are dealing with type II or more special. At the same time we have generally $IJ\neq 0$ so it cannot be just type III. Additionally, the spinor covariant $R_{ABCDEF}$ \cite{Zakhary} has nonzero components
\begin{eqnarray}
	R_{000000}&=&\Psi_{1}(3\Psi_{0}\Psi_{2}-2\Psi_{1}^{2})\,,\\
	R_{000001}&=&\frac{1}{2}\Psi_{2}(3\Psi_{0}\Psi_{2}-2\Psi_{1}^{2})\,,
\end{eqnarray}
which means that generally the spacetime cannot be just of type D. So indeed our scalar field solution is of the algebraic type II which is the most general one in the case of the vacuum Kundt subclass.

Next, we will consider the Ricci tensor whose nonzero frame components are the following
\begin{eqnarray} 
\Phi_{11}&=&\frac{1}{2\,\zeta^{\frac{5}{2}}\,U^2}\,,\\
\Phi_{00}&=&32\,v^2\,\Phi_{11}^2-8\,v\,\Phi_{11}\,\left[h+(\ln{U})_{,u}\right]+\frac{{\phi_{,u}}^2}{2}\,,\\
\Phi_{01}&=&{\frac{\sqrt{2}\Phi_{11}(x-i\,y)}{\sqrt{x^2+y^2}}}\left(U\zeta^{\frac{5}{4}}\left[h+(\ln{U})_{,u}\right]-\frac{4v}{U\zeta^{\frac{5}{4}}}\right)\,,\nonumber\\ \\
\Phi_{02}&=&-2\,\Phi_{11}\,\frac{(y+i\,x)^2}{x^2+y^2}\,,
\end{eqnarray}
hence the Pleba\'{n}ski spinor has three nonzero components $\chi_{0},\chi_{1},\chi_{2}$ and so the Petrov--Pleba\'{n}ski type is II according to the classification process described in \cite{Zakhary2}. In this case the Segre type is necessarily $[2,11]$. This means that the Ricci tensor is non-degenerate and there is no invariance group associated with it. 

From the Weyl scalars and the metric functions one can observe that point $\zeta=0$ (see (\ref{singularity})) looks like a curvature singularity of our solution. We can confirm this by computing the Ricci and the Kretschman scalars
\begin{equation}
  R=\frac{4}{{\zeta}^{5/2}U^{2}}\ ,\quad {\mathcal{K}}=\frac{7}{4}R^{2}\,.
\end{equation}
The singularity is located on the coordinate cylinder with nonzero radius $\rho=\sqrt{x^{2}+y^{2}}=\exp(-\frac{a}{2})$ which lies along the direction of propagation of the possible gravitational waves. This singularity is evidently sourced by the scalar field, namely its spatial part $\psi$ which influences the geometrically important function $\tilde{P}$ through (\ref{equationP}). Considering the metric (\ref{ourmetric}) and the function $\tilde{P}$ (\ref{functionP}) the physical circumference of the cylinder is zero while it is in the finite spatial distance from any point outside of this cylinder. Naturally, one considers only the range of coordinates $x,y$ covering the exterior of this singular cylinder which however (thanks to the vanishing circumference) physically resembles a linear singularity which extends in the $\partial_{v}$ direction.

\section{Scalar waves}\label{previous}
Now we will study the case of a potentially nonvanishing cosmological constant $\Lambda\neq0$. As we have seen in section \ref{first section}, considering the separation of variables (\ref{separation}) leads to vanishing of cosmological constant unless $V_{1}=0=V_{2}$ (see the argumentation culminating after equation (\ref{V1yV2x})). So from now on we are considering $V_{1}=0=V_{2}$ which means that the scalar field is independent of $x,y$ ($\psi=0$ $\Rightarrow$ $\varphi=\varphi(u)$) due to equations (\ref{V1-condition}). Equipped with these information one immediately concludes from (\ref{uu equation}) that the Gaussian curvature of transversal two-spaces is given by the cosmological constant
\begin{equation}
	\Delta(\,\ln P)=\Lambda
\end{equation}
corresponding to the geometrically simple case of constant curvature transversal two-spaces.

The line element (\ref{ourmetric}) now reduces to the following form 
\begin{eqnarray}\label{special case}
\d s^2&=&-H(u,v,x,y)\,\d u^2-2\,\d u\d v +\\
&&+\frac{{U(u)}^2}{\tilde{P}(x,y)^2}(\d x^2+\d y^2)\,.\nonumber
\end{eqnarray}
The Ricci scalar is given by 
\begin{equation}
\mathcal{R}=\frac{2\Delta(\,\ln {\tilde{P}}(x,y))}{U(u)^2}-\frac{\partial^2 H(u,v,x,y)}{\partial v^2}\,,
\end{equation}
As derived above, the scalar field is necessarily only a function of $u$ and the only nonzero component of the energy momentum tensor for such a scalar field is
\begin{eqnarray} \label{energy-momentum component}
{}^{\rm SF}T^v{}_u=-\left({\frac{\partial \varphi(u)}{\partial u}}\right)^2.
\end{eqnarray}
Note that the gradient of the scalar field is now aligned with the null congruence defining the properties of spacetime ($\nabla^{\mu}\varphi \propto \partial_{v}$) which is not possible in the case of Robinson-Trautman family \cite{Tahamtan} where the nonzero expansion of the congruence disallows completely aligned scalar field. Such alignment means that scalar field propagates along this null direction and can be interpreted as a scalar wave. 

From the Einstein tensor components (\ref{xu}),(\ref{yu}) for the metric (\ref{special case}) and the form of the scalar field we obtain $M_{,x}=0=M_{,y}$. This leads to a simplified form of $H$ (see (\ref{definition of H}))
\begin{equation}\label{form of H}
H(u,v,x,y)=-\Lambda\, v^2+h(u)v+K(u,x,y)\,.
\end{equation}
Notice, that for $V_{1}=0=V_{2}$ the quadratic term is automatically fixed from the equation (\ref{xx and yy}). With this form of the metric function $H$ (\ref{form of H}), we obtain two independent Einstein equations with a nontrivial right-hand side (all other equations are already satisfied). The first one corresponds to the general equation (\ref{uu-comp}) and the second one is $G^v{}_u={}^{\rm SF}T^v{}_u$
\begin{eqnarray}
\frac{\Delta(\,\ln {\tilde{P}}(x,y))}{U(u)^2}=\Lambda \,, \label{p equation} 
 \\
\frac{1}{2}\frac{\Delta K}{U(u)^2}-\frac {H_{,v} U'(u)+2U''(u)}{U(u)} =\varphi_{,u}^{2}\,,\label{vu}
\end{eqnarray}
from equation (\ref{vu}) we necessarily have
\begin{equation}\label{K equation}
\Delta K(u,x,y)=C_0\,k(u)\,,
\end{equation}
with some unknown function $k(u)$ and a constant $C_{0}$.
The equation (\ref{p equation}) can only be satisfied if either the nominator and the denominator on its left-hand side are constants or the cosmological constant is vanishing. Thus we will split the investigation into two cases: in the first one we assume $\Lambda=0$ and in the second one we demand $U(u)=const.$ and $\Delta(\,\ln {\tilde{P}})=const.$. 

\subsection{Case $\Lambda=0$}\label{caseA}
In this case it is obvious from (\ref{p equation}) that 
\[\Delta(\,\ln {\tilde{P}}(x,y))=0\,,\]
so the transversal two-spaces spanned by $x,y$ are flat since the above expression is their Gaussian curvature. The function $H(u,v,x,y)$ reduces to
\begin{equation}\label{H0-function}
H(u,v,x,y)=h(u)v+K(u,x,y)\,.
\end{equation}
The relation (\ref{vu}) is the only remaining equation that needs to be solved. Using the equation (\ref{K equation}) the $vu$ component of the Einstein equations (\ref{vu}) now reduces into
\begin{equation} \label{equation1}
\frac{k(u)C_0}{2U(u)^2} -\frac {h(u) U'(u)+2U''(u)}{U(u)} -\left({\frac{\partial \varphi(u)}{\partial u}}\right)^2=0\,,
\end{equation}
and one can generate explicit solutions for $U(u)$ by arbitrarily specifying functions $k(u)$, $h(u)$ and $\varphi(u)$, and a constant $C_{0}$.
We can make the problem even more explicit by assuming the following form of the arbitrary functions
\begin{eqnarray}
h(u)&=&\frac{D(u)}{U'(u)}\,,\\
k(u)&=&{q(u)}{U(u)}\,,\\
\left({\frac{\partial \varphi(u)}{\partial u}}\right)^2&=&\frac{\Phi(u)^2}{U(u)}\,,\label{def-phi}
\end{eqnarray}
and we reduce the problem of solving (\ref{equation1}) to a straightforward double integration with given functions $D, q$ and $\Phi$
\begin{equation}\label{solution-U}
U(u)=\frac{1}{2}\int\int \left(\frac{1}{2}C_0q-\Phi^2-D\right)\d u \d u+C_1u+C_2\,.
\end{equation}
Note, that using (\ref{def-phi}) and (\ref{energy-momentum component}) we conclude that $\Phi^{2}$ is proportional to the energy of the scalar field.

Now we would like to see how general is this family of solutions. Therefore we will determine its algebraic type. The only nonzero Weyl scalar is 
\begin{eqnarray} \label{type1}
\Psi_0&=&\frac{P}{4} \left\{2i(P\,K_{,xy}+P_{,x}K_{,y}+P_{,y}K_{,x})+P\,K_{,yy} \right.\nonumber \\
&&\left.-P\,K_{,xx}+2P_{,y}K_{,y}-2P_{,x}K_{,x} \right\}  \,.
\end{eqnarray}
Now, we can again determine the type irrespective of possible non-optimal choice of tetrad by using explicit methods in \cite{Zakhary}. We immediately see that $IJ=0$ so it is a geometry of the algebraic type III or more special. Additionally, the spinor covariant $Q_{ABCD}$ \cite{Penrose} whose components are quadratic expressions in Weyl scalars is identically zero which means that the algebraic type is N. This is usually interpreted as a geometry representing an exact gravitational wave. The scalar field produces an energy momentum tensor which can be classified according to its Petrov-Pleba\'{n}ski and Segre types (for a review of classification strategies see \cite{Zakhary2}). In our subclass we have the only nonzero component of the tracefree Ricci spinor $\phi_{00}=1/2\varphi_{,u}^{2}$ hence the Pleba\'{n}ski tensor is vanishing and we have Petrov-Pleba\'{n}ski type O. Since $\phi_{11}^{2}-\phi_{01}\phi_{21}=0$ the Segre type is $[(2,11)]$. This Segre type corresponds to radiative sources (a pure radiation field or a null Maxwell field \cite{Stephanietal:book}). The complete solution can then be interpreted as an aligned gravitational and a scalar waves. 

\sloppy
From the equation (\ref{equation1}) one can see that even when the scalar field is vanishing the function $k$ can be nonzero due to the second term of the equation. That means the function $K$ is nontrivial as well (see (\ref{H0-function}) and (\ref{K equation})) and generally the Weyl scalar $\Psi_{0}$ (\ref{type1}) is nonvanishing. Thus even in the absence of any scalar field the gravitational wave can be present. On the other hand, we can as well have a solution with only a scalar wave which is not accompanied by a gravitational one ($\Psi_{0}=0$ $\Rightarrow$ $K$ independent of $x,y$ $\Rightarrow$ $k=0$ from (\ref{K equation})) as can be observed from (\ref{solution-U}) where viable solutions with $\varphi_{,u}\neq 0$ exist for $k=0\Rightarrow q=0$.

\fussy

Note that all type N Kundt spacetimes with null radiation source are known \cite{Stephanietal:book}. However, the above explicit solution represents the source in the form of scalar field which obeys the corresponding field equations. 

We give several explicit solutions based on equation (\ref{equation1}) for specific choices of $k(u)$, $h(u)$ and $\varphi(u)$ in the Table \ref{table1}.

\begin{table*}[t]
$$
\begin{array}{||c|c|c||}
\hline \hline
\varphi(u) & h(u) & U(u) \\
\hline \hline
\gamma \ln{(\frac{\alpha+u}{\beta+u})} &h_{0} & \begin{tabular}[c]{@{}c@{}}$\frac{{(\alpha+u)}^{\frac{1+\epsilon}{2}}}{\e^{\frac{u}{4}(h_{0}+\omega)}}\left\{C_{1}( u+\beta)^{\frac{1+\epsilon}{2}} {\rm HeunC}\left(\frac{\alpha-\beta}{2}\omega,\epsilon,\epsilon,0,\frac{\epsilon^2}{2},\frac{\beta+u}{\beta-\alpha}\right)\right. $\\ $\qquad\qquad\left.\qquad\qquad+ C_{2}( u+\beta)^{\frac{1-\epsilon}{2}} {\rm HeunC}\left(\frac{\alpha-\beta}{2}\omega,-\epsilon,\epsilon,0,\frac{\epsilon^2}{2},\frac{\beta+u}{\beta-\alpha}\right)\right\}$ \end{tabular} \\
\hline
\gamma \ln{(\alpha+\frac{\beta}{u})} & \frac {h_{0}}{u} & \begin{tabular}[c]{@{}c@{}}${\frac{(\alpha u+\beta)^{\frac{1+\epsilon}{2}}}{\e^{\frac{\sqrt{C_{0}}}{2}u}}}\left\{C_{1} u^{ -\frac {h_{0}}{4}+ \frac {1+\Omega}{2}}{\rm  HeunC}\left(\frac{\beta\sqrt{C_{0}}}{\alpha},\Omega, \epsilon,0,\frac{\epsilon^2}{2},-\frac{\alpha}{\beta}u\right) \right. $\\ $\qquad\qquad\left.\qquad\qquad+C_{2}u^{ -\frac {h_{0}}{4}+ \frac {1-\Omega}{2}} {\rm  HeunC}\left(\frac{\beta\sqrt{C_{0}}}{\alpha},-\Omega, \epsilon,0,\frac{\epsilon^2}{2},-\frac{\alpha}{\beta}u\right)\right\}$ \end{tabular} \\
\hline
\alpha u+\beta & h_{0} & \begin{tabular}[c]{@{}c@{}}$ e^{\frac{-u}{4}}\left\{C_{1} \exp{\left(-\sqrt{\omega^2-8\alpha^2}+h_{0}\right)}+C_{2}  \exp{\left(\sqrt{\omega^2-8\alpha^2}+h_{0}\right)} \right\}$ \end{tabular}\\
\hline
\e^{-\alpha u+\beta} & h_{0} & \begin{tabular}[c]{@{}c@{}}$\e^{-\frac{h_{0}u}{4}}\left\{C_{1} {\rm BesselJ}\left(-\frac{\omega}{4\alpha},\frac{1}{\sqrt{2}}\e^{-\alpha u+\beta}\right)\right. $\\ $\quad\qquad\qquad\left.+C_{2} {\rm BesselY}\left(-\frac{\omega}{4\alpha},\frac{1}{\sqrt{2}}\e^{-\alpha u+\beta}\right)\right\}$ \end{tabular}\\
\hline
\sin{\alpha u} & h_{0} & \begin{tabular}[c]{@{}c@{}}$\e^{-\frac{h_{0}u}{4}}\left\{C_{1} {\rm MathieuC}\left(-\frac{\omega^2-4\alpha^{2}}{16\alpha^{2}},-\frac{1}{8},\alpha u\right)\right. $\\ $\quad\qquad\qquad\left.+C_{2} {\rm MathieuS}\left(-\frac{\omega^2-4\alpha^{2}}{16\alpha^{2}},-\frac{1}{8},\alpha u\right)\right\}$ \end{tabular}\\
\hline
\end{array}
$$
\caption{Explicit solutions for $k(u)=U^{2}(u)$, where we have defined $\Omega=\frac{1}{2}\sqrt{-8\gamma^2+h_{0}^2-4h_{0}+4},\ \omega=\sqrt{h_{0}^{2}+4C_{0}},\ \epsilon=\sqrt{-2\gamma^2+1}$}\label{table1}
\end{table*}

\subsection{Case $U(u)=const$ and $\Lambda \neq 0$ }\label{caseB}
In this case we choose $U(u)=1$ for simplicity. One can immediately see from (\ref{p equation}) that 
\begin{equation}\label{expression for lambda}
\Delta(\,\ln {\tilde{P}}(x,y))=\Lambda\,,
\end{equation}
so the transversal two-spaces spanned by $x,y$ have a constant positive or negative curvature based on the value of a cosmological constant. Equation (\ref{expression for lambda}) can be solved explicitly to give $\tilde{P}=1+\frac{1}{4}\Lambda (x^{2}+y^{2})$.

Considering the function $H$ in the form (\ref{form of H}) we can simplify (\ref{vu}) in the following way
\begin{equation}\label{phi-Lambda}
\frac{k(u)C_0}{2} -\left({\frac{\partial \varphi(u)}{\partial u}}\right)^2=0\,,
\end{equation} 
while the function $h(u)$ is unconstrained. The solution is thus specified by providing functions $h, \varphi$ and constants $\Lambda, C_{0}$.

We again compute Weyl scalars for a natural tetrad of this solution and obtain the following nonzero components
\begin{eqnarray} \label{type2}
\Psi_0&=&\frac{\tilde{P}}{4} \left\{2i(\tilde{P}\,K_{,xy}+\tilde{P}_{,x}K_{,y}+\tilde{P}_{,y}K_{,x})+\tilde{P}\,K_{,yy} \right.\nonumber \\
&&\qquad\left.-\tilde{P}\,K_{,xx}+2\tilde{P}_{,y}K_{,y}-2\tilde{P}_{,x}K_{,x} \right\}\,,\\
\Psi_2&=&\frac{\Lambda}{6}-\frac{1}{12}H_{,vv}\,.
\end{eqnarray}
Using \cite{Zakhary,Penrose} we can immediately confirm that $I^{3}=27J^{2}$ is satisfied so that we are dealing with type II or more special. At the same time we have generally $IJ\neq 0$ so it cannot be just type III. Additionally, the spinor covariant $R_{ABCDEF}$ has a nonzero component
\begin{eqnarray}
	R_{000001}&=&\frac{1}{2}\Psi_{2}(3\Psi_{0}\Psi_{2}-2\Psi_{1}^{2})\,,
\end{eqnarray}
which means that generally the spacetime cannot be just of type D. So indeed our scalar field solution is of the algebraic type II which is the most general one in the case of vacuum Kundt subclass. We have one nonzero component of the tracefree Ricci spinor $\phi_{00}=\frac{1}{4}\Delta H$ hence the Petrov-Pleba\'{n}ski type is O and the Segre type is $[(2,11)]$ indicating again that the source corresponds to a null radiation given by the scalar wave.

The above described case is a generalization of specific subcase of Kundt solutions with a cosmological constant described in \cite{Lewandowski}. Namely, our solution additionally supports the scalar field.

If one would want to restrict the algebraic type solely to D (keeping $\Psi_{2}\neq 0$ but setting $\Psi_{0}=0$) one can derive from (\ref{type2}) that necessarily $K$ is independent of $x,y$ which by using (\ref{K equation}) means $k=0$. But that leads (see (\ref{phi-Lambda})) to a vanishing contribution of the scalar field energy momentum tensor. In accordance with \cite{GriffithsPodolsky:book} we might interpret our type II (here we refer to algebraic type of the Weyl tensor) solution for the metric as an exact gravitational wave on a type D background. As we have just seen this (nonradiative) background cannot support nonvanishing scalar field. So the scalar wave is necessarily accompanied by the gravitational wave in a kind of nontrivial interaction where the scalar wave generates the gravitational one. Unlike in the previous case of a vanishing cosmological constant \ref{caseA} where both waves could exist independently. This type D background is a direct product spacetime, specifically the Nariai ($dS_{2}\times S^{2}$) or anti-Nariai ($AdS_{2}\times H^{2}$) solution based on the value of the cosmological constant.

Of particular interest might be a question whether our subclasses admit pp-wave solutions that still retain the scalar field. Due to the construction of the Kundt class the geometry reduces to that of the pp-wave if the principal null direction ${\bf l}=\partial v$ is covariantly constant
\begin{equation}\label{covariantly-constant}
	l_{\alpha;\beta}\,\d x^{\alpha}\d x^{\beta}=-\frac{1}{2}\frac{\partial H(u,v,x,y)}{\partial v}=0\ .
\end{equation}
For the case of the model described in subsection \ref{caseA} this translates into $h=0$ and $D=0$. In the case of model described in subsection \ref{caseB} we have $h=0$ and $\Lambda=0$. In both cases the scalar field is generally nonvanishing so we have pp-wave solutions with a scalar field.

\section{Nonlinear Electrodynamics}
So far we have considered only a scalar field possibly with a nonvanishing cosmological constant. It is interesting to investigate the possibility of having some form of the nonlinear electrodynamics (NE) as an additional source. We assume the Lagrangian of the nonlinear electromagnetic field $\mathcal{L}(F)$ to be an arbitrary function of the invariant $F=F_{\mu \nu}F^{\mu \nu}$ constructed from a closed Maxwell 2-form $F_{\mu \nu}$ with the following energy momentum tensor  
\begin{equation}\label{energy-momentum-Maxwell}
{}^{\rm NE}T^{\mu}{}_{\nu}=\frac{1}{2}\{\delta^{\mu}{}_{\nu} \mathcal{L}-(F_{\nu \lambda}F^{\mu \lambda})\mathcal{L}_F\}\,,
\end{equation}
which contributes to the right-hand side of the Einstein equations
\begin{equation}\label{NE field equations}
G^{\mu}{}_{\nu}={}^{\rm SF}T^{\mu}{}_{\nu}+{}^{\rm NE}T^{\mu}{}_{\nu}-\Lambda \delta ^{\mu}{}_{\nu}\,,
\end{equation}
while the modified Maxwell (nonlinear electrodynamics) field equations are given in the following form 
\begin{equation} \label{modified Maxwell}
\partial _{\mu}(\sqrt{-g}\mathcal{L}_FF^{\mu \nu})=0\,,
\end{equation} 
in which $\mathcal{L}_F=\frac{d\mathcal{L}(F)}{dF}$. We are specifically interested in adding the NE source to the solution derived in the section \ref{caseB}. Then the Maxwell 2-form aligned with the principal null direction has necessarily the following form (considering (\ref{energy-momentum-Maxwell}) and (\ref{special case}))
\begin{equation} 
\mathbf{F}=E(u,v,x,y) \d u \wedge \d v \,,
\end{equation} 
then from (\ref{modified Maxwell}) and the metric (\ref{special case}) one can find 
\begin{equation} \label{NEQ}
 \mathcal{L}_F F_{uv}=F_{0}(x,y)\,,
\end{equation} 
where $F_{0}$ is an arbitrary function. However, since the electromagnetic field invariant can be expressed as $F=-2F_{uv}^{2}$ one immediately concludes (considering the functional dependence in (\ref{NEQ})) that actually $F_{0}$ is necessarily a constant and $E(u,v,x,y)=E(u,v)$. The energy momentum tensor given in (\ref{energy-momentum-Maxwell}) can be expressed in the form 
\begin{equation}\label{energy-momentum-NE}
 {}^{\rm NE}T^{\mu}{}_{\nu}=diag\left\{\frac{\mathcal{L}}{2}-F\mathcal{L}_F,\frac{\mathcal{L}}{2}-F\mathcal{L}_F,\frac{\mathcal{L}}{2},\frac{\mathcal{L}}{2}\right\}\,.
\end{equation}
As one can see from (\ref{NEQ}) with $F_{0}=const$, it is possible to find the form of nonlinear electrodynamics Lagrangian explicitly 
\begin{equation} \label{form of L}
\mathcal{L}(F)=-\alpha \sqrt{-F}\,,
\end{equation}
where $\alpha=\sqrt{8}F_{0}$. This result is consistent with the $uu$ component of (\ref{NE field equations}) as well (when considering $U(u)=const.$)
\begin{equation}
-\Delta(\,\ln {\tilde{P}}(x,y))+\Lambda=\frac{\mathcal{L}}{2}-F\mathcal{L}_F\,.
\end{equation}
It is clear that $\frac{\mathcal{L}}{2}-F\mathcal{L}_F$ should be a constant. For simplicity one may choose this constant as zero. Then the form of NE Lagrangian would be the same as (\ref{form of L}) and we will satisfy (\ref{expression for lambda}) as well. 

This particular form of Lagrangian, namely a square root of invariant $F$, is not new and it has been investigated previously (see e.g. \cite{Guendelman,square root}). Namely, it was shown that the spherically symmetric purely electric solution is absent in this model, or, in other words, the electric monopole is vanishing by definition. On the other hand, gauge theory with such a Lagrangian contains interesting string-like solutions \cite{Aurilia} leading to possible confinement. Also, radiation modes (or null electromagnetic field solutions satisfying $F=0$) do not appear naturally in this model but can be recovered using a magnetic condensation in effective four-dimensional theory coming from a 6D compactification. This square root Lagrangian is also a special case of the so called Power Maxwell model \cite{power Maxwell} ($(-F)^{s}$) when $s=\frac{1}{2}$ which was a subject of intensive study. One problem in this type of models is usually connected with the energy conditions, however, we can easily avoid this by properly selecting the constant $\alpha$.  

The next equation is the $xx$ component of (\ref{NE field equations})
\begin{equation}\label{xx}
\frac{1}{2}\frac{\partial^2 H(u,v,x,y)}{\partial v^2}+\Lambda=\frac{\mathcal{L}}{2} \,.
\end{equation}
If we define 
\begin{equation}\label{Hdef}
H(u,v,x,y)=\chi(u,v)-\Lambda v^2+h(u)v+K(u,x,y)\,,
\end{equation}
the above equation reduces to a simpler form
\begin{equation} \label {Lagrangian in terms of uv}
\mathcal{L} =\frac{\partial^2 \chi(u,v)}{\partial v^2}\,.
\end{equation}
Since in NE the (mixed component) energy momentum tensor does not have any non-diagonal terms the $vu$ component of (\ref{NE field equations}) has the scalar field as the only source. So we can retrieve the equation which is the same as (\ref{phi-Lambda}).

The electromagnetic field considered above is evidently non-null ($F_{\mu\nu}F^{\mu\nu}\neq 0$) and since $F_{\mu\nu}{}^{*}\hspace{-0.07cm}F^{\mu\nu}=0$ it is purely electric in a preferred frame. The Petrov type of the Weyl tensor is easily determined since the nonzero Weyl scalars are still given by (\ref{type2}) and so the type is II. The complete energy momentum tensor (containing both (\ref{energy-momentum-NE}) and (\ref{energy-momentum component})) is of Petrov--Plena\'{n}ski type D since we have nonzero tracefree Ricci spinor components $\phi_{00}=\frac{1}{4}\Delta H, \phi_{11}=\frac{1}{8}\chi_{,vv}$ and the Segre type is then $[2,(11)]$ according to classification scheme in \cite{Zakhary2}.

One may be surprised that we have non-null electromagnetic field but there is no source in the nonlinear Maxwell equation (\ref{modified Maxwell}). However, using Leibniz rule one can split the left-hand side of this equation and rearrange it into the following form 
\[
\partial _{\mu}(\sqrt{-g} F^{\mu \nu})=-\sqrt{-g} F^{\mu \nu}\partial_{\mu}\ln{\mathcal{L}_F}\,.
\]
Now, the left-hand side represents the standard Maxwell equation and the right-hand side represents a source generated by nontrivial $\partial_{\mu}\ln{\mathcal{L}_F}$. So from the point of view of the standard Maxwell theory a solution of the vacuum nonlinear electrodynamics has an effective source. This interpretation was already noted in the original paper \cite{BornInfeld}.

Finally we give the explicit form of $E(u,v)$. From (\ref{Lagrangian in terms of uv}) and the relation $F=-2E^2$ we immediately get
\begin{equation}\label{E-field}
|E(u,v)|=-\frac{1}{4F_{0}}\frac{\partial^2 \chi(u,v)}{\partial v^2} \,.
\end{equation}

Here we can again search for a pp-wave geometry in our results. In this case, from equations (\ref{Hdef}) and (\ref{covariantly-constant}) we see that the Lagrangian (\ref{Lagrangian in terms of uv}) and the electric field (\ref{E-field}) itself necessarily vanishes. However, if we do not impose the definition (\ref{Hdef}) and analyze equation (\ref{xx}) directly we can see that the Lagrangian and therefore the electric field is constant. So the pp-wave spacetime in our class of solutions generally admits only a constant electric field together with a scalar field. 

\section{Conclusion and Final Remarks}
We have investigated a free massless scalar field coupled to the Kundt spacetime. The explicit solutions were given based on the behavior of Gaussian curvature of transversal two-spaces. The first explicit solution corresponds to a spacetime with singular coordinate cylinder with the singularity sourced by the scalar field. Physically this singularity behaves just as a linear singularity due to the vanishing circumference of the cylinder. This singularity lies along the principal null direction which is at the same time the direction of propagation of gravitational waves that are present for $\Psi_{0}\neq 0$ (see (\ref{Weyl-singular})). This solution can be thought of as an analogue of the scalar field solution in the Robinson--Trautman class given in \cite{Tahamtan}.

We have also given two explicit (up to simple integration) subclasses of the Kundt family containing scalar field wave. Their respective algebraic types of the Weyl tensor are N and II. The notable absence of type D  solution with a scalar field confirms the results of \cite{Bergh} where it was shown that Kundt type D subclass does not admit null massless scalar field. One can easily see that our assumption about the functional dependence of the scalar field necessarily means that it is null. This scalar field is naturally radiative. Since our type II Kundt geometry (corresponding to the case with $\Lambda\neq 0$) is interpreted as an exact gravitational wave on type D background we can see that the presence of a scalar wave necessarily generates accompanying gravitational wave. On the other hand, in the type N solution (corresponding to the case with $\Lambda= 0$) the gravitational and scalar waves can exist independently. So on the level of our Kundt spacetime with a scalar field the interdependence of the waves of both fields is determined by the presence of a cosmological constant. Both classes of spacetimes admit pp-wave solutions with scalar field as a special case.

The scalar wave solution with vanishing cosmological constant (section \ref{caseA}) satisfies the conditions of being a VSI spacetime as considered in \cite{Coley-PRL}. Namely, it possesses a privileged null congruence ($\mathbf{n}=\partial_{v}$) which is geodesic and all its optical scalars are vanishing, the Petrov type is N and the Ricci tensor is aligned as well $R_{\mu\nu}\sim n_{\mu}n_{\nu}$ leading to the Petrov--Pleba\'{n}ski type O. Then, according to \cite{Coley-PRL}, all higher order corrections to this classical solution coming from string theory ($\sigma$-model) perturbations necessarily vanish and we have a solution valid in the string theory as well (plus theories with higher order modifications to Einstein--Hilbert action). This result concerns possible modification of the metric coming from the string theory, but we have a scalar field present as well. Our scalar field is not of the dilatonic type considered in \cite{Coley-PRL} which is naturally appearing in string theory. However, one can similarly argue that all potential higher-order corrections which would be given in terms of second-rank tensors and scalars constructed from the metric, $\nabla_{\mu}\varphi$ and their derivatives must vanish. So that this class of solutions solves a broad range of theories that not only contain modifications to the equations governing the geometry but also the scalar field. A special class of these solutions are generalised pp-waves with a scalar field which were considered in the realm of the string theory already in \cite{Horowitz} using argumentation similar to \cite{Coley-PRL}. 

In the final section we have considered a general electrodynamic field with nonlinear Lagrangian as an additional source and obtained solutions for the geometry, the electromagnetic field and the specific form of Lagrangian. The physical significance of the derived Lagrangian was investigated in several previous works. In this case the pp-wave condition results in trivial solution containing only constant electric field. The algebraic type of the Weyl tensor was II in this case. The electromagnetic field is non-null and sourceless. However, it has a source generated by the nonlinear Lagrangian when interpreted in the scope of the Maxwell theory. 

\begin{acknowledgements}
We would like to thank J. Bi\v{c}\'{a}k for enlightening discussions. This work was supported by the grant GA\v{C}R 17-13525S. 
\end{acknowledgements}

\section*{Appendix}
Let us briefly consider the possibility of having a complex scalar field $\Phi$ with the energy momentum tensor given by
\[
T_{\mu \nu}=\frac{1}{2}(\nabla_{\mu}\Phi^{*} \nabla_{\nu}\Phi+\nabla_{\mu}\Phi\, \nabla_{\nu}\Phi^{*}-g_{\mu \nu}\,g^{\alpha\beta}\nabla_{\alpha}\Phi \,\nabla_{\beta}\Phi^{*})
\]
which evidently still satisfies the null energy condition. One of the ways how to avoid the problem of inheritance (the matter fields inheriting the geometrical properties of the underlying spacetime) is to prepare the scalar field in a specific form that leads to certain subtractions due to complex conjugation. In the case of our scalar field and the problem of retaining the dependence on the coordinate $v$ the setup would be 
\[
\Phi=\e^{iV(v)}\,\varphi(u,x,y)
\]
\sloppy
One can easily check that the off-diagonal components $T_{vx},T_{vy}$ are vanishing now although they would be present for a real scalar field dependent on all coordinates. 

We can check now if this idea helps with retaining the $v$-dependence in our case. The independence of the scalar field on $v$ was derived in equation (\ref{v-constraint}) which translates into condition $T_{vv}=0$. However the complex scalar field $\Phi$ provides
\[
T_{vv}=\left(\frac{\partial V}{\partial v}\right)^{2}\varphi^{2}
\]
again leading to $v$-independence.

\fussy

The possible route how to include the dependence on the coordinate $v$ might be a ghost scalar field (having negative kinetic term in Lagrangian) which violates the null energy condition. This means that the form of the metric (\ref{ourmetric}) as described in \cite{Stephanietal:book,GriffithsPodolsky:book}) is no longer a general one for nonexpanding and nontwisting class of solutions. Namely, the potential presence of shear might provide an additional freedom needed for the nontrivial $v$-dependence.

\end{document}